\documentclass[10pt,twocolumn,amsmath,amssymb,floatfix,showpacs]{revtex4}
\usepackage{graphicx,epsfig,latexsym,amssymb}
\usepackage{multirow,amsmath,array,booktabs}
\usepackage{dcolumn}
\usepackage{color}
\usepackage{textcomp}
\usepackage{longtable}
\usepackage[section]{placeins}
\usepackage{bm}

\begin{document}

\title{Self-consistent description
of the halo nature of $^{31}$Ne with continuum and pairing correlations}

\author{S. S. Zhang$^{1}$}\email{zss76@buaa.edu.cn}
\author{B. Shao$^{1}$}
\author{S. Y. Zhong$^{1}$}
\author{M. S. Smith$^{2}$}\email{smithms@ornl.gov}
\affiliation{$^{1}$School of Physics, Beihang University, Beijing 100191, China}
\affiliation{$^{2}$Physics Division, Oak Ridge National Laboratory, Oak Ridge, Tennessee, 37831-6354, USA}

\begin{abstract}

\textbf{Background:}
A relativistic structure model has previously been used to predict a halo structure for $^{31}$Ne [S. S. Zhang, M. Smith, Z. S. Kang and J. Zhao, Phys. Lett. B {\bf 730}, 30 (2014)], consistent with halo signatures from measured reaction cross sections of Ne isotopes bombarding Carbon targets.
However, previous attempts to calculate those cross sections with reaction models were missing contributions from resonances and pairing correlations in their structure input.

\textbf{Purpose:}
Use a reaction model with our relativistic fully microscopic structure model input to predict these cross sections and momentum distributions and analyze for possible halo signatures.

\textbf{Methods:}
Structure input for exotic Ne isotopes were obtained via the analytical continuation of the coupling constant (ACCC) method based on the relativistic mean field (RMF) theory with Bardeen-Cooper-Schrieffer (BCS) pairing approximation, the RAB approach. Total reaction cross sections, one-neutron removal cross sections, and momentum distributions of breakup reaction products were calculated with a Glauber model using our relativistic structure input.

\textbf{Results:}
Our predictions of total reaction and one-neutron removal cross sections of $^{31}$Ne on a Carbon target were significantly enhanced compared with those of neighboring Neon isotopes, agreeing well with measurements at 240 MeV/nucleon and consistent with a single neutron halo. Furthermore, our calculations of the inclusive longitudinal momentum distribution of the $^{30}$Ne and valence neutron residues from the $^{31}$Ne breakup reaction indicate a dilute density distribution in coordinate space, another halo signature.

\textbf{Conclusions:}
We give a full description of the halo nature of $^{31}$Ne that includes a self-consistent use of pairing and continuum contributions that makes predictions consistent with reaction cross section measurements. This approach can be utilized to determine the halo structure of other exotic nuclei.

\end{abstract}

\maketitle

\newpage
\section{Introduction}
Nuclei are considered to have a ``halo'' structure when they have a valence nucleon (or nucleons) with an extended spatial distribution, with more than half of the valence probability density lying outside the range of the potential of the remaining core nucleons~\cite{Khalili2004,Jensen2004,Hansen1987}. The weak binding of the valence nucleon to the core via the short range nuclear force enables it to tunnel outward to large (classically forbidden) radial distances~\cite{Khalili2004}, likely as a single-particle resonant state in the low-energy continuum ~\cite{Ring1996}. The nuclear halo is a threshold effect~\cite{Jensen2004}, resulting in a low-density valence nucleon distribution surrounding a dense nuclear core, very different from standard nuclei. This leads to the display of interesting effects (\textit{e.g.}, clustering~\cite{Wurzer1997}, influencing the neutron dripline~\cite{Jensen1992,Nakamura2009}), but also makes halo nuclei a challenging test for nuclear models.

While the shell model cannot reproduce halo structure, numerous theoretical structure models have been used to describe halo nuclei, including the No-Core Shell Model~\cite{Navratil2002}, Greens Function Monte Carlo~\cite{Pieper2001}, Coupled Cluster~\cite{Dean2003}, and relativistic Hartree Bogoliubov methods~\cite{Meng1996, Zhou2010}. These last two (and similar) studies emphasized the critical role that nuclear pairing plays in halo nuclei via the coupling nucleon pairs in bound states with those in the continuum. Furthermore, it has long been realized that halos cannot be formed with any angular momentum value, but only for $s$- and $p$-waves~\cite{Sagawa1992,Riisager1992}.

The first indication that the exotic neutron-rich nucleus $^{31}$Ne may have a halo structure came from a measurement of its large Coulomb breakup cross section on Pb and C targets at the RIBF facility at RIKEN~\cite{Nakamura2009}. As the first $p$-orbital single neutron halo structure, and the heaviest halo nuclei at the time, the mechanism of halo formation in $^{31}$Ne has been of significant interest. In early 2014, a relativistic structure model was successfully used to predict a halo structure for $^{31}$Ne. The analytical continuation of the coupling constant (ACCC) method, based on a relativistic mean field (RMF) theory with the Bardeen-Cooper-Schrieffer (BCS) pairing approximation -- the RAB model~\cite{ZhangPLB14} -- was used to calculate neutron- and matter-radii, one-neutron separation energies, $p$- and $f$-orbital energies and occupation probabilities, and neutron densities for single-particle resonant orbitals in $^{27-31}$Ne. These results were analyzed for evidence of neutron halo formation in $^{31}$Ne due to a competition of the occupation for unbound resonant orbitals and the pairing correlations. Based on a radius increase from $^{30}$Ne to $^{31}$Ne, that is much larger than the increase from $^{29}$Ne to $^{30}$Ne,
and a simultaneous decrease in the neutron separation energy, a $p$-orbital 1$n$ halo structure was predicted for $^{31}$Ne.
Later that year, the spin and parity of its valence neutrons were confirmed from experiment~\cite{Nakamura2014},
consistent with our predictions, and finally $^{31}$Ne was determined to have a $p$-orbital halo structure.

Matter radii, neutron density distributions, and occupation probabilities are, however,
not \textit{observables} for halo nuclei. Evidence of halo structure has been overwhelmingly based on total reaction cross sections, Coulomb breakup cross sections, nucleon removal cross sections, and momentum distributions following nuclear breakup~\cite{Khalili2004}. While semi-classical
(\textit{e.g.}, Glauber-like~\cite{Glauber59, Abu-Ibrahim2003}) cross section models usually work well at the high energies of these measurements, attempts to explain the trends (\textit{i.e.}, large increase) in the measured cross section of $^{31}$Ne bombarding a $^{12}$C target with self-consistent theoretical reaction models, without any additional effects, have so far been nearly or out of 1$\sigma$ agreement with measured data. For example, structure input from a microscopic self-consistent model within non-relativistic mean field framework, \textit{i.e.}, a Skyrme Hartree Fock (SHF) model, were used~\cite{Horiuchi2012} as input for a Glauber model, but underestimated the measured large increase in the interaction cross section at $^{31}$Ne (compared to $^{30}$Ne), regardless of the force (\textit {e.g.}, SLy4 or SkM*) used in the model. This may be due to lack of contributions from pairing correlations and the continuum in the SHF model.
A different approach utilizing Antisymmetrized Molecular Dynamics (AMD) combined with a double-folding model largely underestimated the reaction cross section for $^{31}$Ne + $^{12}$C $\to$ $^{30}$Ne + X. However, by further adding a Resonating Group Method (RGM), issues of the density ``tail'' were partially improved, enabling predictions to be nearly 1$\sigma$ below the measured data~\cite{Takenori2012}.

To pursue a consistent structure and reaction modeling approach to describe $^{31}$Ne,
we use a Glauber reaction model~\cite{Abu-Ibrahim2003} with structure input from
the relativistic, fully microscopic model~\cite{ZhangPLB14} that has already successfully predicted the halo structure nature of $^{31}$Ne.
Our models are briefly described in Section II, followed in Section III by our predictions of total reaction cross sections, one-neutron removal cross sections, and momentum distributions of Ne isotopes bombarding a $^{12}$C target. We compare these predictions to measurements and analyze for reaction halo signatures in $^{31}$Ne, and then summarize in Section IV.

\section{Formulism and Numerical Details}
The Glauber model~\cite{Glauber59,Abu-Ibrahim2003} takes nucleon-nucleon interactions and nuclear densities as input to calculate total, reaction, and scattering cross sections in a semi-classical formulism.
It involves an integral over the impact parameter of the nuclear transmission,
which contains the density distributions of the target and projectile~\cite{Khalili2004}.
Since this model has no free parameters~\cite{Mehndiratta2017},
it is widely used in many applications, such as the determination of the radii for halo nuclei from measured cross sections (\textit{e.g.},~\cite{Tanihata1985,Horiuchi2007}).


In this study, we use the formulism for single-nucleon halo nuclei as given in Ref.~\cite{Abu-Ibrahim2003}.
The equations for the reaction cross section of projectile-target and core-target are defined by~\cite{Abu-Ibrahim2003}:
\begin{equation}
	\sigma _{reac}(P+T)=
	\int (1-|\langle\psi _0|e^{i\chi _{CT}(\textbf{b}_C)+i\chi _{NT}
	(\textbf{b}_C+\textbf{s})}|\psi _0\rangle|^2)\,d\textbf{b},
\label{Eq1}
\end{equation}
\begin{equation}
	\sigma _{reac}(C+T)=\int (1-|e^{i\chi _{CT}(\textbf{b})}|^2)\,d\textbf{b},
\label{Eq2}
\end{equation}
where $\psi_0$ is the valence-nucleon wave function, $\textbf{b}_C$  denotes the impact parameter between core and target,
and \textbf{s} refers to the two-dimensional coordinates comprising the $x$- and $y$-components of coordinate of projectile (or target) to its center of mass coordinate.
The nucleon-target phase-shift function, $\chi_{NT}$, and the core-target phase-shift function, $\chi_{CT}$,
are defined through the relevant densities
\begin{equation}
	i\chi_{CT}(\textbf{b})=-\int d\textbf{r}\int d\textbf{r}'\rho _C(\textbf{r})\rho _T(\textbf{r}')\Gamma(\textbf{b}+\textbf{s}-\textbf{s}'),
\label{Eq3}
\end{equation}
\begin{equation}
	i\chi_{NT}(\textbf{b})=-\int d\textbf{r}\rho_T(\textbf{r})\Gamma(\textbf{b}-\textbf{s}).
\label{Eq4}
\end{equation}
Here the density $\rho$ is normalized by $\int \rho(\textbf{r})\,d\textbf{r}=A$.
And the profile function $\Gamma$ is defined as the following~\cite{Abu-Ibrahim2003}:
\begin{equation}
\Gamma(\textbf{b})=\frac{1-i\alpha}{4\pi\beta}\sigma_{NN}e^{(-\textbf{b}^2/2\beta)},	
\label{Eq5}
\end{equation}
\noindent where the parameters $\sigma_{NN}, \alpha, \beta$ are defined in Ref.~\cite{Abu-Ibrahim2008}.
The connection between structure and reaction models occurs via the nuclear density distributions,
which are obtained from a structure model and as the input into the reaction model.

One-neutron removal cross section $\sigma_{-{N}}$ quantifies the process
whereby the projectile nucleus breaks up into a core and one neutron in a continuum state.
This cross section is approximately equal to the difference between the projectile-target and core-target reaction cross sections,
i.e.
\begin{equation}
	\sigma_{-N}=\sigma _{reac}(P+T) - \sigma _{reac}(C+T).
\label{Eq6}
\end{equation}

%
\begin{figure}
\includegraphics[scale=0.23]{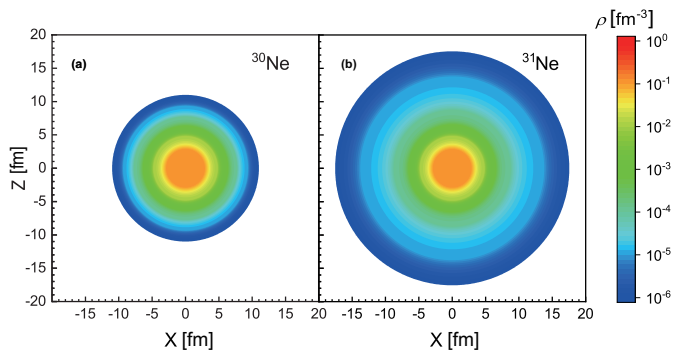}
\caption{Density distribution in spherical nuclei $^{30}$Ne (a) and $^{31}$Ne (b).}
\label{Fig1}
\end{figure}
In our previous study~\cite{ZhangPLB14}, we used the RAB approach to describe the structure of each of the exotic $^{27 - 31}$Ne nuclei as a core plus a single valence neutron in a resonant orbital.
Since the density distribution of $^{30}$Ne and $^{31}$Ne are crucial input in Glauber model, we show them in Fig.~\ref{Fig1}.
It can be clearly seen that the density distribution of $^{31}$Ne has a larger spatial dilution than that of $^{30}$Ne,
which is consistent with a halo structure in $^{31}$Ne.
\begin{table}[h]
\caption{The RMSDs $\sigma$ for the number $N$ of GFs.}
\begin{ruledtabular}
\begin{tabular}{ccccc}
$N$&4&5&10&15\\
	\hline
$\sigma$ (fm$^{-3}$)&$2.9\times10^{-4}$ & $1.9\times10^{-4}$&$1.4\times10^{-4}$ &$1.5\times10^{-4}$  \\
\end{tabular}
\end{ruledtabular}
\label{Tab1}
\end{table}
For simplicity, the density distribution of the core nucleus for the projectile 
and that of the target are fitted by a {\it set} of $N$ Gaussian functions (GFs)~\cite{Abu-Ibrahim2003},
\begin{equation}
\rho(r)=\sum_i^N c_i \exp(-a_i r^2),
\label{Eq7}
\end{equation}
with parameters $a_i$  and $c_i$.
Meanwhile, the valence-neutron wave function $\psi_0$ can be directly read from the output of the RAB model.
In this way, the core-target and nucleon-target phase-shift functions can be analytically expressed,
which result in obtaining the reaction cross section in an analytical way~\cite{Karol1974}.

\begin{figure}[ht]
\includegraphics[scale=0.55]{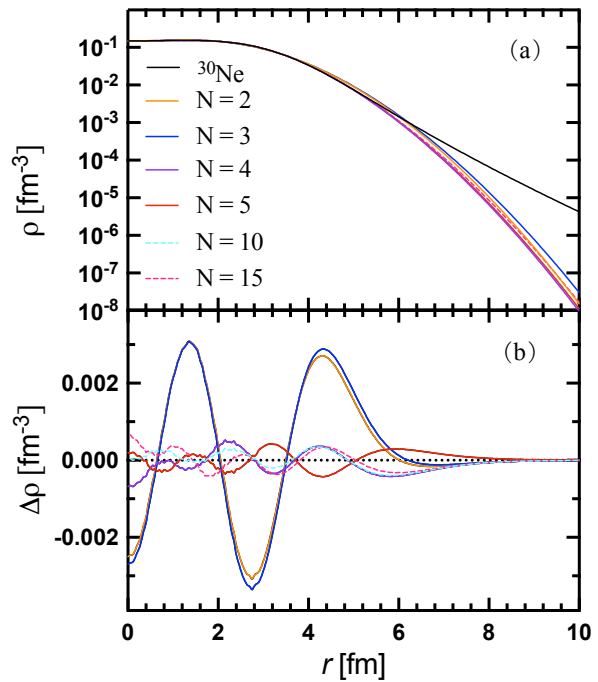}
\caption{(a) Our fitting densities (colored lines) with different number $N$ of GFs to that
for the core nucleus $^{30}$Ne (black line) and (b) the deviations $\Delta\rho$ as a function of radius $r$.}
\label{Fig2}
\end{figure}

For neon isotopes, the core density $\rho(r)$ in $^{26, 28, 30}$Ne from the RAB model are fit as in Eq.~(\ref{Eq7}).
Taking the core $^{30}$Ne nucleus as an example, we study the behaviour of fitting GFs with the set number $N$ increasing
and plot the deviation $\Delta\rho(r)$ from the exact RAB density in Fig.~\ref{Fig2}(a) and Fig.~\ref{Fig2}(b), respectively.
It can be seen that our fittings can be improved with the increase of $N$,
\textit{e.g.} from $N = 2$ to $N = 5$.
Nevertheless, we need to point out that over-fitting may appear for larger $N$, for example $N=20$.
From Fig.~\ref{Fig2}(b), we can see that the deviations mainly come from the range of radius less than 8 fm,
and the corresponding deviations are less than $\pm 5 \times 10^{-4}$ fm$^{-3}$.

We also evaluate the root mean square deviation (RMSD) $\sigma$ of fitting GFs
from the exact density in the range of $ r \le $ 16 fm with a step of 0.05 fm,
which is defined by
\begin{equation}
\sigma = \sqrt{\dfrac{\sum_{i=1}^{N'}\Delta\rho^2}{N'}}.
\label{Eq8}
\end{equation}
For chosen $N$, the number of GFs, we list the RMSD results in TABLE~\ref{Tab1} for comparison.
It is clear that proper fittings can be reached when $N=4$ or $5$, which results in the RMSD $\sigma \sim$ $10^{-4}$ fm$^{-3}$.
Accordingly, the reaction cross section $\sigma_R$ using the Glauber model is approximately 1453 mb.
When $N=2$ and $N=3$, the fitting results with the RMSD $\sigma$ $\sim 1.2*10^{-3}$ fm$^{-3}$,
will cause the deviation of the above $\sigma_R$ by 2.5$\%$.
The RMSD $\sigma$ is reduced to  $\sim$ $10^{-4}$ fm$^{-3}$ when the number of GFs $N$ increases to 4 or 5 or larger.
However,  this results in a change in $\sigma_{R}$ by less than $3$ \textperthousand,
while significantly increasing the time to calculate the momentum distribution. For these reasons,
we use $N=5$ GFs to fit the density of the core nucleus $^{30}$Ne in the following calculations.
By fitting the density for the core nucleus and the wave function for the valence neutron from the RAB model, 
we can compute the nucleon-target and the core-target phase- shift functions using Eqs.~(\ref{Eq3}) and (\ref{Eq4}). 
Then, the reaction cross section of projectile-target and core-target can be obtained by Eqs.~(\ref{Eq1}) and (\ref{Eq2}).
Sudden increase of the reaction cross sections for the isotopes or large one-neutron removal cross section of the neighboring nuclei,
are both signatures of a halo structure.

Another crucial quantity to directly determine a halo structure is
the inclusive longitudinal momentum distribution of the core nucleus and the valence neutron residues from the breakup reaction.
The longitudinal momentum distribution of the core fragment after the inelastic breakup of the projectile is as follows~\cite{Abu-Ibrahim2003}:
\begin{equation}
\begin{split}
\frac{d\sigma^{inel}_{-N}}{dP_{\left|\right|}}=\frac{1}{2\pi\hbar}\int(1-e^{-2Im\chi_{NT}(\textbf{b}_N)})\,d\textbf{b}_N\\
\times\int e^{-2Im\chi_{CT}(\textbf{b}_N-\textbf{s})}\,d\textbf{s}\\
\times\int\,dz\int e^{\frac{i}{\hbar}P_{\left|\right|}(z-z')}u^*_{nlj}(r')u_{nlj}(r)\frac{1}{4\pi}P_l(\hat{\textbf{r}}',\hat{\textbf{r}})\,dz',
\end{split}
\label{Eq9}
\end{equation}
\noindent where $\textbf{r}=(\textbf{s}, z)$ and $\textbf{r}'=(\textbf{s}, z')$.
After inelastic breakup of the projectile into a core plus a continuum neutron, a narrow longitudinal momentum distribution
implies a large spatial distribution of the projectile, which is a halo signature.

\section{Results and Discussion}
Using the Glauber model, regarding Gaussian-fitted RAB-based core density and the wave function for valence neutron from RAB model as the input,
we compute the total reaction cross sections of the exotic $^{26 - 31}$Ne isotopes bombarding a $^{12}$C target,
one-neutron removal cross section and the longitudinal momentum distribution by Eqs. (\ref{Eq1}), (\ref{Eq2}), (\ref{Eq6}), (\ref{Eq9}), respectively.
We search for signatures of a halo utilizing a self-consistent description of the nuclear structure and nuclear reactions of these nuclei.
\begin{figure}
\includegraphics[scale=0.80]{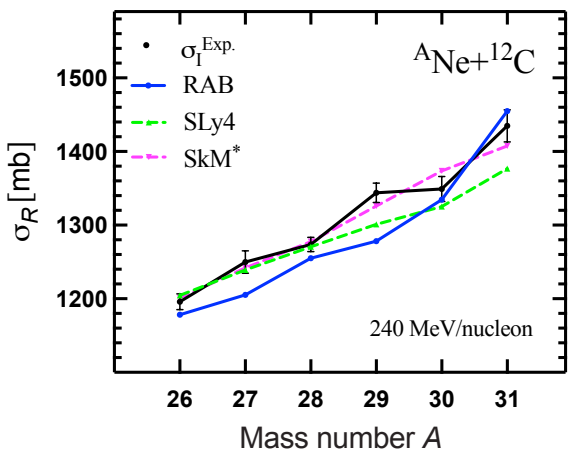}
\caption{Reaction cross section $\sigma_R$ of~$^{26 - 31}$Ne bombarding a $^{12}$C target at 240 MeV/nucleon versus mass number $A$. Shown are experimental data from Ref.~\cite{Takechi2010}, current Glauber model predictions with RAB structure input, and previous SHF theory results (denoted by SLy4 and SkM$^*$) from Ref.~\cite{Horiuchi2012}.}
\label{Fig3}
\end{figure}

Fig.~\ref{Fig3} shows our Glauber model reaction cross section calculations $\sigma_R$ for  $^{26 - 31}$Ne bombarding a $^{12}$C target at 240 MeV/nucleon. The measured interaction cross sections $\sigma_I$ ~\cite{Takechi2010} are also shown in this figure; since the difference between $\sigma_I$ and $\sigma_R$ is small at 240 MeV/nucleon, we do not distinguish them for comparison. Looking at the trends in this plot, we see that the measured cross section increases by an average of $39 \pm 7$ mb/amu from $^{26}$Ne to $^{30}$Ne, which is matched by our predicted increase of 39 mb/amu. The unit mb/amu refers to the change of reaction cross section for 1 unit change of atomic mass.
Our values are within 3$\sigma$ of the measured values, except for $^{29}$Ne;
this will be the topic of a separate study in the near future.
Our focus here, however, is on reaction signatures of a halo structure in $^{31}$Ne. Experimentally, that signature is the large increase in the measured cross section from $^{30}$Ne to $^{31}$Ne, by $86 \pm 39$ mb/amu, up by a factor of $2.2 \pm 1.1$ from the average change over the range $^{26}$Ne to $^{30}$Ne.
Our prediction also shows a large matching increase in cross section from $^{30}$Ne to $^{31}$Ne compared with the average increase from $^{26}$Ne to $^{30}$Ne, by a factor of 3, clearly a signature of an increase in radius characteristic of a halo nucleus -- and one that is consistent with our RAB model calculations of the structure of $^{31}$Ne.

Fig.~\ref{Fig3} also shows a previous attempt in Ref.~\cite{Horiuchi2012} to match the cross section trends using a Glauber model with non-relativistic SHF structure model input. Their calculation with the SkM$^{*}$ force, while producing values that agree within 2$\sigma$ of the experimental values, has a trend that disagrees with measurements: the cross section slope from $^{30}$Ne to $^{31}$Ne is 20\% \textit{lower} than the slope from $^{26}$Ne to $^{30}$Ne. This trend certainly is inconsistent with a halo signature. Their calculation with the SLy4 force, while exhibiting a slope increase of 80\% from $^{30}$Ne to $^{31}$Ne, would need to have more than twice that slope increase in order to match the measured cross section at $^{31}$Ne. Neither calculation can match the cross section values and trend (large increases) that signify a halo signature of $^{31}$Ne. This may be a consequence of a lack of contributions from pairing correlations and the continuum,
in the SHF model, aspects which play a crucial role in the formation of the halo structure in our RAB model. In order to verify this conjecture, we use the $^{30}$Ne density from Ref.~\cite{Horiuchi2012} with our valence nucleon wave function to calculate the reaction cross section in the Glauber model. The resulting calculated cross section is approximately 1435 mb,
an increase of 5.7\% over the value of SLy4 result and 3.3\% over the value of SkM* result in Ref.~\cite{Horiuchi2012} and in much better agreement with the experimental data. This confirms the importance of contributions from pairing correlations and the continuum.

\begin{figure}
\includegraphics[scale=0.80]{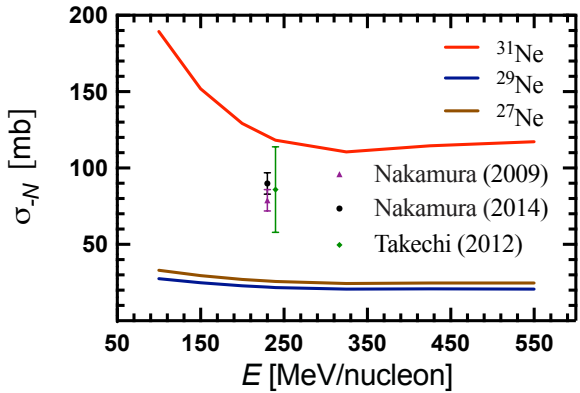}
\caption{One-neutron removal cross section $\sigma_{-N}$ of $^{27,29,31}$Ne on a $^{12}$C target as a function of incident energy $E$. 
The experimental data are from Refs.~\cite{Nakamura2009, Nakamura2014, 2012Takechi}.}
\label{Fig4}
\end{figure}

An examination of our predicted one-neutron removal cross sections $\sigma_{-N}$ for the exotic Ne isotopes $^{27,29,31}$Ne, Fig.~\ref{Fig4}, further shows reaction-based evidence for a halo structure via Glauber model calculations with RAB structure input. Our calculated $^{31}$Ne 1$n$ removal cross section is approximately a factor of 6 times larger than that of neighboring $^{29}$Ne and $^{27}$Ne, clearly a halo signature for $^{31}$Ne. Our results are consistent with measured values for $^{31}$Ne~\cite{Nakamura2009, Nakamura2014, 2012Takechi} that are enhanced by a factor of 3 - 6 times than those of the lower-mass nuclides. We also note that our predictions have an energy dependence that agrees with the latest measurements. The 2014 remeasurement of the 230 MeV/nucleon cross section~\cite{Nakamura2014} has a larger value than their first measurement~\cite{Nakamura2009}, which demonstrates a decreasing cross section from 230 MeV/nucleon to the 240 MeV/nucleon measurement of Ref.~\cite{2012Takechi}. We note, however, higher precision measurements in this range are needed to firmly determine the cross section energy dependence. Although the calculated $\sigma_{-N}$ for $^{31}$Ne is not in exact agreement with the experimental data, it is significantly higher than those of adjacent isotopes, indicating the halo structure of $^{31}$Ne.

\begin{figure}
\includegraphics[scale=0.80]{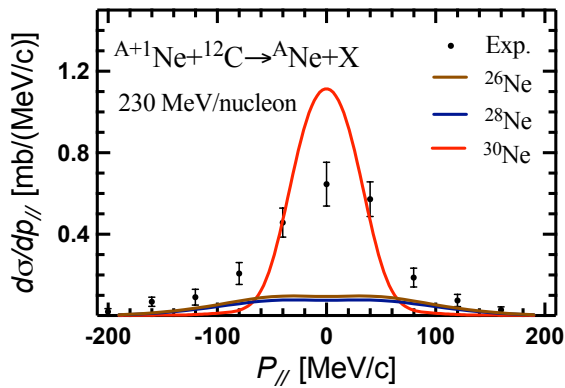}
\caption{Inclusive parallel momentum distribution of $^{26,28,30}$Ne residues after 1$n$ removal from $^{27,29,31}$Ne bombarding a $^{12}$C target.
The experimental data are taken from Ref.~\cite{Nakamura2014}.}
\label{Fig5}
\end{figure}

We use Glauber model to calculate momentum components of the (post-breakup) core that are parallel to the projectile axis, and examine the overlap with the (plane-wave) continuum neutron. The inclusive longitudinal momentum distributions that we calculate will be examined for trends and compared to measurements.
Fig.~\ref{Fig5} shows one additional piece of reaction-based halo evidence in our calculations for $^{31}$Ne, this based on the inclusive longitudinal momentum distribution after inelastic breakup of the projectile into a core plus a continuum neutron.
Our calculated distributions for $^{30,28,26}$Ne residues are shown after one-neutron removal from $^{31,29,27}$Ne bombarding a $^{12}$C target at 230 MeV/nucleon, respectively. We predict a narrowly peaked momentum distribution for $^{30}$Ne, with a full width at half maximum (FWHM) of approximately 76 MeV/c, in comparison to the much flatter distributions for $^{28,26}$Ne. This is consistent with a dilute spatial distribution in the $^{31}$Ne projectile before breakup in accordance with the Heisenberg uncertainty principle of quantum mechanics, a signature of halo nuclei.
Also shown in Fig.~\ref{Fig5} is a measurement of the $^{30}$Ne momentum distribution~\cite{Nakamura2014}, which also shows a peaked distribution but with a larger FWHM (approximately $120\pm 10 $ MeV/c). We do note that our calculated momentum distribution is lower than the measurements near 100 MeV/c. This may be due to the small contribution from the $f$-orbital in the RAB model~\cite{ZhangPLB14}. Furthermore, our overall calculated distribution may be modified if we consider the effects of deformation.

\section{Summary}
A description of the halo nature of $^{31}$Ne from the perspective of both structure and reactions is realized for the first time. We used the relativistic, fully microscopic RAB structure model, which previously predicted a $p$-orbital single neutron halo structure for $^{31}$Ne based on a large increase in radius, to generate core and valence nucleon densities as input for a Glauber reaction model. Our calculations for $^{31}$Ne bombarding a $^{12}$C target at 240 MeV/nucleon led to three reaction-based halo nuclei signatures.

Firstly, our increase of reaction cross section from $^{30}$Ne to $^{31}$Ne is 3 times larger than those of adjacent Ne isotopes, which is consistent with measurements. Our calculated one-neutron removal cross sections were a factor of 6 higher than neighboring nuclei and decreasing in energy, again consistent with measurements.

Secondly, we calculate a narrow longitudinal momentum distribution of inelastic breakup products consistent with a dilute density distribution in coordinate space for projectile $^{31}$Ne. But neighboring nuclei $^{27,29}$Ne show much flatter momentum distributions. Our momentum distribution was qualitatively similar to, but not as wide as, the measured distribution. The combination of the RAB structure model and the Glauber reaction model have now predicted all of these halo signatures for $^{31}$Ne: large neutron- and matter-radii, enhanced one-neutron separation energies, high $p$-orbital occupation probabilities, enhanced total reaction cross section, enhanced one-neutron removal cross section, and narrow breakup residue momentum distribution.

We anticipate that this self-consistent structure and reaction model approach will be very useful to explain other exotic halo nuclei in the future. Further studies, including deformed structures and related work, are in progress.

\section{Acknowledgements}
{
This work was partially supported by the
National Natural Science Foundation of China under Grant
No.~11375022, No.~11775014, and by the U.S. Department of Energy Office of Science, Office of Nuclear Physics, under Award Number DE-AC05-00OR22725.}

\end{document}